\documentclass[12pt]{article}
\usepackage{authblk}
\usepackage{graphicx} 
\usepackage{amssymb}
\usepackage{amsmath}
\usepackage{amsthm}

\usepackage{tikz}
\usepackage{tikz-cd}

\usepackage{xcolor}
\usepackage{color}
\definecolor{darkgreen}{rgb}{0,0.5,0}
\usepackage[colorlinks=true, linkcolor=darkgreen, urlcolor=blue, citecolor=red]{hyperref}

\title{Ruijsenaars duality for \(B, C, D\) Toda chains}

\begin{small}
\author[1]{Ivan Sechin}
\affil[1]{\small Beijing Institute of Mathematical Sciences and Applications, \protect\\
Huairou district, Beijing 101408, China, 
\href{mailto:sechin@bimsa.cn}{sechin@bimsa.cn}}

\author[2]{Mikhail Vasilev}
\affil[2]{\small School of Mathematics and Statistics, University of Glasgow, \protect\\
University Place, Glasgow G12 8QQ, UK, 
\href{mailto:Mikhail.Vasilev@glasgow.ac.uk}{Mikhail.Vasilev@glasgow.ac.uk}}
\end{small}
\date{}

\begin{document}

\maketitle

\begin{abstract}
We use the Hamiltonian reduction method to construct the Ruijsenaars dual systems to
generalized Toda chains associated with the classical Lie algebras of types
\(B, C, D\). The dual systems turn out to be the $B, C$ and $D$ analogues of 
the rational Goldfish model, which is, as in the type $A$ case, the strong coupling limit 
of rational Ruijsenaars systems. We explain how both types of systems emerge
in the reduction of the cotangent bundle of a Lie group and provide the formulae
for dual Hamiltonians. We compute explicitly the higher
Hamiltonians of Goldfish models using the Cauchy--Binet theorem.    
\end{abstract}

\tableofcontents

\section{Introduction}
Toda systems \cite{RS, R} are well-known beautiful examples of finite-dimensional integrable models
having deep connections with many different areas of mathematics and physics such as representation theory
\cite{E, K1}, supersymmetric gauge theories \cite{GKMMM, G1}, quantum cohomology \cite{GK}, matrix models
\cite{GMMMO} and many others. Integrable Toda chains were introduced by M. Toda in \cite{T}.
Since then they have become a prominent example of many-body integrable systems. 
In \cite{OP, P} generalizations of certain integrable models related to root systems were introduced. 
In this paper, we will deal with open non-relativistic Toda systems for classical root 
systems $B, C$, and $D$. Our main aim is to establish Ruijsenaars duality between Toda systems and the so-called 
rational Goldfish models using the methods of Hamiltonian reduction. 
This duality is a certain relationship between the
classical integrable many-body systems, which connect action variables of one of the systems with
the coordinates of the dual one, and vice versa. This duality was first introduced as the relation
between models in the Calogero--Ruijsenaars family, where it is very well understood,
except for elliptic models \cite{BMMM}.

However, the duality in the Toda case was not so well studied in the classical case beyond the simplest
type $A$ case. It is well-known that the Toda family could be considered as a scaling
(Inozemtsev) limit of Calogero--Ruijsenaars systems \cite{I}. 
On the one hand, we know that the trigonometric Calogero--Moser--Sutherland model is dual to the 
rational Ruijsenaars system, on the other hand, we know that open non-relativistic Toda chains could 
be obtained via the Inozemtsev limit from the trigonometric Calogero--Moser--Sutherland model, while
the rational Goldfish model could be obtained as a strong coupling limit of a rational
Ruijsenaars system. 
It turns out that these limiting procedures are in agreement with Ruijsenaars duality in the sense
that starting from two dual systems and performing these limits we arrive at two dual systems again.
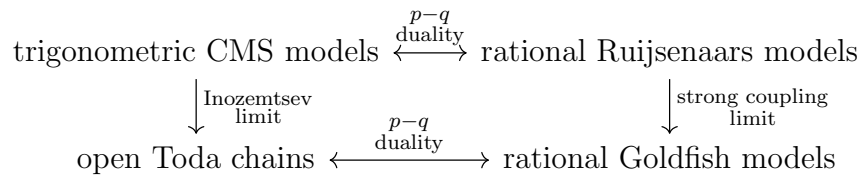
\begin{equation*}
\begin{tikzcd}
    \text{trigonometric CMS models}
        \arrow[r, leftrightarrow, "\substack{p-q \\ \text{duality}}"]
        \arrow[d, "\substack{\text{Inozemtsev} \\ \text{limit}}"] &
    \text{rational Ruijsenaars models} 
        \arrow[d, "\substack{\text{strong coupling} \\ \text{limit}}"] \\
    \text{open Toda chains} 
        \arrow[r, leftrightarrow, "\substack{p-q \\ \text{duality}}"] &
    \text{rational Goldfish models}
\end{tikzcd}
\end{equation*} 
Let us note that there exists quite a vast number of interrelations between
the dualities in integrable systems and other areas \cite{MM, FGNR, Mars}. 

Let us give a brief introduction to the Goldfish models introduced by Calogero \cite{C}. These 
systems appear as the strong coupling limit of Ruijsenaars systems, with a coupling parameter
that tends either to \(0\) or \(\infty\). For instance, the trigonometric Goldfish model
corresponds to $t = 0$ classical Ruijsenaars--Macdonald operators. In the same manner, 
the rational Goldfish model considered here is a degeneration of the rational 
Ruijsenaars system. The rational and trigonometric Goldfish systems are also connected with
the probabilistic integrable systems of TASEP type in the spirit of quantum-classical duality
\cite{GVZ}, and also appear as examples of the integrable systems obtained via the induced dynamics
approach \cite{P}. Both rational and trigonometric Goldfish models could also be derived 
from the dynamics of poles of the Novikov--Veselov equation \cite{Z}. 

Let us note that there also exists a quantum version of Ruijsenaars duality \cite{R4}, 
usually called bispectrality, which states that for two quantum dual systems, there exists a joint 
family of eigenfunctions, where the first system acts on the coordinate variables and the second
system acts on spectral variables. The eigenfunctions for (q)-Toda chains are called
(q)-Whittaker functions \cite{K, K2, K1, E}, and the bispectrality property for Whittaker functions
was extensively studied in the literature \cite{B, SK, DE}.

In this paper, we further develop the ideas of L. Feh\'er and many of his collaborators, who extensively
studied the Ruijsenaars duality using the methods of symplectic reduction 
\cite{FA, FG, FK, FK1, FK2, FK3, FMarshal, FP}. In this paper, we focused on generalizing the results 
of \cite{F} to the case of classical root systems other than $A$. Our approach is quite 
straightforward and simple, we start with the cotangent bundle to the classical complex simple 
Lie groups $SO(2n,\mathbb{C})$, $SO(2n + 1,\mathbb{C})$, $Sp(2n,\mathbb{C})$ with the natural
actions of unipotent and maximal compact subgroups together with two families of invariant 
Poisson-commuting Hamiltonians on this cotangent bundle. We then calculate the Hamiltonian 
reduction in two different ways to obtain two Ruijsenaars dual systems, which result in
\(B, C, D\) generalized open non-relativistic Toda chains and rational Goldfish models.

\section{Ruijsenaars duality in geometric setting}
\subsection{Ruijsenaars duality}
Ruijsenaars duality, also well-known as $p-q$ or action-angle duality, is a certain relationship
between two classical many-body integrable systems. We call two systems dual to each other in 
the Ruijsenaars sense if there exists a (anti)symplectomorphism between the underlying phase 
spaces which maps the coordinates of the first system to the action variables of the second 
and vice versa. This duality was introduced by S. Ruijsenaars in his series of papers \cite{R1, R2, R3}.

The first and simplest example of this duality is the self-duality of the rational Calogero--Moser
system, which was already present in the paper \cite{KKS}. The next non-trivial example
is the duality between the trigonometric Calogero--Moser system and the rational Ruijsenaars model,
which could be lifted to self-duality for the trigonometric Ruijsenaars system, which reflects on the
classical level symmetry in the Macdonald theory. For an elementary and 
transparent review of Ruijsenaars duality, we refer to a recent report \cite{ZLB}.

The symplectic reduction procedure provides a powerful method for constructing Ruijsenaars dual
systems \cite{MW}. In a nutshell, we start with a big phase space with a group-theoretical origin 
(usually cotangent bundles to Lie groups or their analogues), with a natural Hamiltonian action of
a Lie group. On this big phase space, there exist two different sets of invariant Poisson-commuting 
Hamiltonians. 
When calculating the reduction two different sections of the corresponding principal bundle could 
be used to describe the reduction explicitly. For one of the sections the first family of the 
Hamiltonians become trivial consisting of coordinates of the integrable system whose integrals 
are given by the second family of Hamiltonians, and the same for the second section. Thus, we
arrive at two, in principle different, integrable systems, which are, however, connected 
via Ruijsenaars duality due to our knowledge of the reduction procedure.

\subsection{Reduction scheme}
Let \(G\) be a simple matrix complex Lie group, \(B_+\), \(B_{-}\) --- its opposite Borel
subgroups, \(N_+, N_-\) the corresponding unipotent subgroups, and \(K\) its maximal compact 
subgroup. Denote by \(\mathfrak{g}\) the Lie algebra of \(G\), 
\(\mathfrak{n}_+\) and \(\mathfrak{k}\) --- Lie algebras of \(N_+\) and \(K\) respectively.
Let \(\mathrm{Tr} \colon \mathfrak{g} \times \mathfrak{g} \to \mathbb{C}\) be the 
invariant bilinear form on Lie algebra \(\mathfrak{g}\), this form is used to identify
\(\mathfrak{g}^* \simeq \mathfrak{g}\). Let $N_{-}$ be the opposite unipotent subgroup and 
$\mathfrak{n}_{-}$ -- its Lie algebra, then using the invariant bilinear form we can obtain 
the following identifications $\mathfrak{n}_{-}^{*} \simeq \mathfrak{n}_{+}$ and 
$\mathfrak{k}^{*} \simeq \mathfrak{k}$. 

Consider a cotangent bundle  \(T^* G\) with the canonical symplectic structure. This bundle
can be trivialized by left translations 
\(T^* G \simeq G \times \mathfrak{g}^* = \{(g, X) \mid g \in G, \ X \in \mathfrak{g}^*\}\).
Then the canonical one-form and the corresponding symplectic form are given by
\begin{equation}
    \label{cotangent_bundle_symplectic_form}
    \alpha = \mathrm{Tr}(X g^{-1} dg), \quad
    \omega = d \alpha = 
        \mathrm{Tr}(dX \wedge g^{-1} dg) - 
        \mathrm{Tr}(X g^{-1} dg \wedge g^{-1} dg).
\end{equation}
The group \(G\) acts on itself by left and right actions.
These actions can be lifted to Hamiltonian actions of \(G\) on the cotangent bundle \(T^* G\)
\begin{gather}
    \label{action_cotangentbudnle}
    L_h \colon T^* G \to T^* G, \quad L_h(g, X) = (hg, X), \\ \notag
    R_h \colon T^* G \to T^* G, \quad R_h(g, X) = (gh^{-1}, h X h^{-1}).
\end{gather}
The corresponding momentum maps are
\begin{gather}
    \label{momentum_left_right}
    \mu_L \colon T^* G \to \mathfrak{g}^*, \quad
        \mu_L(g, X) = g X g^{-1}, \\ \notag
    \mu_R \colon T^* G \to \mathfrak{g}^*, \quad
        \mu_R(g, X) = -X.
\end{gather}
We will consider the left action of the complex unipotent subgroup \(N_+\) and the right action of
the maximal compact subgroup \(K\) on \(G\) and \(T^* G\). These actions are  
Hamiltonian with the momentum maps given by the projections on the corresponding Lie
subalgebras
\begin{gather}
    \label{momentum_unipotant_compact}
    \mu_{N_+} \colon T^* G \to \mathfrak{n}_+^* \simeq \mathfrak{n}_-, \quad
        \mu_{N_+}(g, X) = \mathrm{Pr}_{\mathfrak{n}_-}(g X g^{-1}), \\ \notag
    \mu_K \colon T^* G \to \mathfrak{k}^* \simeq \mathfrak{k}, \quad
        \mu_{K}(g, X) = -\mathrm{Pr}_{\mathfrak{k}}(X).
\end{gather}
Now consider the action of \(N_+ \times K\)
\begin{equation}
    N_+ \times K \colon T^* G \to T^* G, \quad
        (g, X) \mapsto (n g k^{-1}, k X k^{-1})
\end{equation}
it has the momentum map
\begin{equation}
    \label{momentum}
    \mu \colon T^* G \to \mathfrak{n}_- \times \mathfrak{k}, \quad
        \mu(g, X) = 
            (\mathrm{Pr}_{\mathfrak{n}_-}(g X g^{-1}),
            -\mathrm{Pr}_{\mathfrak{k}}(X)).
\end{equation}
Let us choose the following value of the momentum map 
\begin{equation}
    \label{value}
    \lambda = \left(\lambda_{\mathfrak{n}_{-}}, \lambda_{\mathfrak{k}} \right) =
        \Big(\sum_{\alpha \in \Pi} e_{-\alpha}, \ 0\Big) \in
            \mathfrak{n}_- \times \mathfrak{k},
\end{equation}
where \(\Pi\) is the set of simple roots of \(\mathfrak{g}\).
Such a choice of the value of the momentum map is because the stabilizer of
\(\lambda\) under the coadjoint action coincides with the whole group
\(\mathrm{Stab}(\lambda) = N_+ \times K\).
According to the Hamiltonian reduction procedure, the reduced space
\begin{equation}
    M_\lambda = \mu^{-1}(\lambda)/(N_+ \times K)
\end{equation}
is a symplectic manifold, because the group action is free and proper. The usual calculations with dimensions show that
\begin{equation*}
    \dim_{\mathbb{R}} M_\lambda = \dim T^*G - 2 \dim N_+ - 2 \dim K = 
    2 \dim G - 2 \dim N_+ - 2 \dim K = 2\ \mathrm{rk}\ G.
\end{equation*}
Let us consider the following diagram
\begin{equation}
\begin{tikzcd}
    \mu^{-1}(\lambda) \arrow{d}{\pi} \arrow[hookrightarrow]{r}{i} & T^* G \\
    M_\lambda &
\end{tikzcd}
\end{equation}
where \(i\) is the inclusion map, and \(\pi\) is the projection map.
According to the principles of Hamiltonian reduction \cite{MW} the symplectic form 
\(\omega_{red}\) on the reduced space \(M_\lambda\) is uniquely defined by the 
following relation
\begin{equation}
    i^* \omega = \pi^* \omega_{red}.
\end{equation}
If \(f, g\) are \(N_+ \times K\) invariant functions on \(T^*G\) they define 
functions \(\tilde{f}, \tilde{g}\) on the reduced phase space \(M_\lambda\)
\begin{equation}
    i^* f = \pi^* \tilde{f}, \quad
    i^* g = \pi^* \tilde{g}.
\end{equation}
The Poisson bracket \(\{f, g\}_{T^* G}\) of two invariant functions is also
an invariant function, then it also defines a function on \(M_\lambda\) such 
that
\begin{equation}
    i^* \{f, g\}_{T^*G} = \pi^* \{\tilde{f}, \tilde{g}\}_{M_\lambda}.
\end{equation}
This fact provides a powerful method of constructing the nontrivial Poisson-commuting
functions on the reduced space \(M_\lambda\), starting from simple invariant 
Poisson-commuting functions on the cotangent bundle \(T^*G\). If we take a sufficient
number of such functions, we obtain a classical integrable system after reduction.

\subsection{Toda gauge and Moser gauge}

In this paper, we will consider only classical complex simple Lie groups of type 
$A_n, B_n, C_n, D_n$ together with the complex reductive group $GL(n,\mathbb{C})$ 
in their vector representations and we will write $X^{\dagger}$ for the Hermitian
conjugate matrix of $X$. For an explicit description of classical root systems and
matrix representation of these Lie algebras, one can consult Appendix 1. 

In these matrix cases, there exist two natural sets
of \(N_+ \times K\) invariant Poisson-commuting functions on the cotangent bundle 
$T^*G$, namely,
\begin{gather}
    \label{I} 
    I_k = \mathrm{Tr}(X^k), \\
    \label{J} 
    J_k = m_k(g g^{\dagger}),
\end{gather}
where $m_k(A)$ denotes the $k \times k$-size minors of the matrix $A$, constructed
from $k$ bottom rows and $k$ right columns of the matrix $A$. These two families of
invariant Poisson-commuting functions provide two families of Poisson-commuting 
Hamiltonians after the reduction. The number of independent Hamiltonians of either
type is $n = \mathrm{rk} \ G$, thus, they provide two integrable systems on the
reduced symplectic manifold. These integrable systems appear to be dual in the sense
of Ruijsenaars (action-angle, $p-q$) duality.

The first approach is to bring the matrix \(g\) to the diagonal form using Iwasawa 
decomposition \cite{Iwasawa} for complex semisimple Lie group. 
Let \(\theta \colon \mathfrak{g} \to \mathfrak{g}\) be the Cartan involution, which
acts as \(\theta \colon X \mapsto -X^\dagger\). Since \(\theta^2 = 1\), it yields 
the eigenspace decomposition \(\mathfrak{g} = \mathfrak{k} \oplus \mathfrak{p}\), 
where \(\mathfrak{k}\) and \(\mathfrak{p}\) correspond to eigenvalues \(+1\) and 
\(-1\) respectively. Let \(i \mathfrak{a}\) be a maximal abelian subalgebra in 
\(\mathfrak{k}\), then the Cartan subalgebra \(\mathfrak{h}\) of \(\mathfrak{g}\) is
\(\mathfrak{h} = \mathfrak{a} \oplus i \mathfrak{a}\). Denote by \(A\) the connected Lie
group corresponding to the Lie algebra \(\mathfrak{a}\). Then every \(g \in G\) 
can be uniquely represented in the form \(g = n a k\), where 
\(n \in N_+, \ a \in A, \ k \in K\). Therefore, the joint action of $N_+ \times K$ 
accomplishes the diagonalization of the matrix $g$
\begin{equation} 
    g = e^q, \quad q = \sum_{i = 1}^{\mathrm{rk} G} q_i h_i, \ q_i \in \mathbb{R}
\end{equation}
We will call this Toda gauge, following L. Feher's notations \cite{F}. 
Resolving the momentum map equation, one
can find the explicit form of the matrix $X$ up to Cartan component $p$
\begin{equation}
    X = p + \sum_{\alpha \in \Pi} e^{(\alpha, q)} (e_{\alpha} + e_{-\alpha}), 
        \quad p = \sum_{i = 1}^{\mathrm{rk} G} p_i h_i, \ p_i \in \mathbb{R}.
\end{equation}
In the Toda gauge the Hamiltonians (\ref{I}) become non-trivial Poisson-commuting 
functions on the reduced space $M_{\lambda}$, which provides us well-known
Poisson-commuting Hamiltonians of the Toda integrable system related to the corresponding
Lie algebra
\begin{equation}
    H_k = \frac{1}{k}\mathrm{Tr}(X^k),
\end{equation}
in particular
\begin{equation}
    H_2 = \sum_{i = 1}^{\mathrm{rk} G} \frac{p_i^2}{2} + 
        \sum_{\alpha \in \Pi} e^{2 (\alpha, q)}.
\end{equation}

The second approach is to diagonalize the matrix \(X\) using the coadjoint action of the
maximal compact subgroup $K$. Note that any \(X\) in the preimage $\mu^{-1}(\lambda)$
has the zero projection onto \(\mathfrak{k}\) due to the second momentum map equation
(\ref{momentum}), thus, the action of \(K\) brings \(X\) to the diagonal form 
\(X = \sum_{i = 1}^{\mathrm{rk} G} \hat{q}_i h_i, \ \hat{q}_i \in \mathbb{R}\). 
However, in this gauge, we have the residual \(N_+\) action,
which enables us to represent the matrix \(g\) uniquely as a matrix from the negative
Borel \(B_-\) subgroup, at least on the dense subset of $G$. We will call this the Moser gauge.
Notice that for \(X \in \mathfrak{h}\) and \(g \in B_-\) the result of the adjoint
action 
\begin{equation}
    \mathrm{Ad}_g(X) = g X g^{-1} \in \mathfrak{b}_- = \mathfrak{h} \oplus \mathfrak{n}_-.
\end{equation}
Moreover, \(\mathrm{Pr}_{\mathfrak{h}}(g X g^{-1}) = X\), which enables us to rewrite the 
first momentum map equation (\ref{momentum}) in the Moser gauge as
\begin{equation}
    g X g^{-1} - X = \lambda_{\mathfrak{n}_{-}}.
\end{equation}
Using this form of the momentum map equation we would be able to obtain a recursive formula
for the matrix elements of $g$.
In this gauge the Hamiltonians (\ref{J}) become non-trivial Poisson-commuting
functions on the reduced space $M_{\lambda}$, and form the Goldfish integrable system \cite{C}.
The direct computations in the Moser gauge will be done separately for each root system
in the following sections.

We would like to emphasize that in \cite{F} another idea was used for calculations
in the Moser gauge, and we use a different approach to generalize it to the cases
of other root systems.

\subsection{Computations of minors}

To write the explicit formulae for the Hamiltonians of the Goldfish model, we need to compute 
the lower-right $k \times k$ minors of a matrix $g g^\dagger$, which is a product of a 
lower-triangular matrix $g$ and its Hermitian conjugate $g^\dagger$. 
The useful tool for this computation is the Cauchy--Binet formula, which allows us
to rewrite the determinant of a product of two rectangular matrices into a sum over the
minors of the multipliers. 

Namely, let \(A\) be an \(k \times m\) matrix and \(B\) be an \(m \times k\) matrix, 
then the determinant of a product \(AB\) can be computed by the following formula
\begin{equation}
    \det(AB) = \sum_{S \in \binom{[m]}{k}} \det(A_{[k], S}) \det(B_{S, [k]}),
\end{equation}
where \(\binom{[m]}{k}\) is the set of all subsets of \(\{1, \ldots, m\}\) of size \(k\).
\(A_{[k], S}\) is a square \(k \times k\) matrix constructed from the columns of \(A\), 
labeled by indices from \(S\), and \(B_{S, [k]}\) is an \(nk\times k\) matrix formed of
the rows of \(B\), labeled by indices from \(S\).

Let us now compute the lower-right minors of \(g g^\dagger\) using this formula.
Notice that these minors can be rewritten as determinants of products of rectangular
matrices
\begin{equation}
    m_k(g g^\dagger) = \det(g_k g_k^\dagger),
\end{equation}
where \(g_k\) is a matrix formed of \(k\) lower rows of the matrix \(g\). Thus, we can
apply the Cauchy--Binet formula, which is simplified in this case to the sum of the squares
of modules of minors of \(g\)
\begin{equation}
    m_k(g g^\dagger) = \sum_{S \in \binom{[k]}{n}} |\det((g_k)_{[k],S})|^2.
\end{equation}

One of our main tools for computing the minors, which are the parts of this sum, is based
on a technical result, suggested by Ruijsenaars \cite{R}. Let \(x = \{x_1, \ldots, x_m\}\),
\(b = \{b_1, \ldots, b_m\}\) and \(M(b, x)\) be an \(m \times m\) matrix of the form
\begin{equation}
\label{M_matrix}
    M_{ij} (b, x) = \begin{cases}
        b_j \cdot \prod\limits_{k = j + 1}^i \dfrac{1}{x_j - x_k}, \quad &j \le i, \\
        0, \quad &j > i.
    \end{cases}
\end{equation}
Then the following result computes the minors of this matrix. Denote by
\(V_{i_1, \ldots , i_k}\) the \(k \times k\) minor composed of
the columns \(i_1 < i_2 < \ldots < i_k\) and last \(k\) rows of matrix $M(b, x)$, then
\begin{equation}
    \label{detV}
    V_{i_1, \ldots , i_k} = 
    \prod_{\substack{t, s = 1 \\ t < s}}^k (x_{i_t} - x_{i_s})
    \prod_{r = 1}^k \prod_{j = i_r + 1}^m \frac{1}{x_{i_r} - x_j} \cdot
    \prod_{l = 1}^k b_{i_l}.
\end{equation}
Note that in this formula the product in the enumerator cancels out with the multipliers in
the denominator and we get the alternative formula 
\begin{equation}
    \label{detV1}
    V_{i_1, \ldots , i_k} = 
    \prod_{r = 1}^k \prod_{\substack{j = i_r + 1 \\ j \notin \{i_1, \ldots, i_k\}}}^m
        \frac{1}{x_{i_r} - x_j} \cdot
    \prod_{l = 1}^k b_{i_l}.    
\end{equation}
In the following sections, this approach will be used to compute all the Hamiltonians
for the Goldfish models associated with root systems $A, B, C, D$.

\section{\(A_{n - 1}\) case}
Remark that we work for convenience with $G = GL(n, \mathbb{C})$ rather than 
$SL(n, \mathbb{C})$. The only difference in the result is that for $SL(n, \mathbb{C})$ 
we work in the center of mass systems and we always need to impose certain conditions 
on variables.
Here we assume that $G = GL(n, \mathbb{C})$, maximal compact subgroup is $K = U(n)$ --- 
a group of unitary matrices and $N_{+}$ --- unipotent group of upper-triangular matrices
with units on the diagonal. In this case, we choose an invariant bilinear form on 
$\mathfrak{gl}(n, \mathbb{C}) = \mathrm{Lie} (GL(n,\mathbb{C}))$ to be the usual 
trace form. The preimage of the momentum map is explicitly written in terms of the 
momentum map equations
\begin{equation}
\label{moment1A}
         \mathrm{Pr}_{\mathfrak{n}_-}(g X g^{-1})_{ij} = \delta_{i-1,j}, 
\end{equation}
\begin{equation}
\label{moment2A}
    \mathrm{Pr}_{\mathfrak{k}}(-X) = 0.
\end{equation}
Equation (\ref{moment2A}) tells us that matrix $X$ is Hermitian. Now we solve the momentum
map equations in two different ways.

\subsection{Toda gauge}
To compute Hamiltonian reduction, we firstly diagonalize matrix $g$ using actions of groups
$N_{+}$ and $U(n)$ via the Iwasawa decomposition to bring $g$ to diagonal form with positive
entries $e^q = e^{\sum_{i = 1}^n q_i h_i} = \mathrm{diag} (e^{q_1}, \ldots, e^{q_n})$, 
where $q_i$ are real.
Then, the equation (\ref{moment1A}) reads as
\begin{equation}
    e^{q_i - q_j} X_{ij} = \delta_{i - 1, j}, \quad i > j,
\end{equation}
so the elements of the matrix $X$ below the diagonal are uniquely determined by these equations.
The upper-diagonal part is recovered if we recall that $X$ is Hermitian. Let us denote 
$X_{ii} = p_i$, thus
\begin{equation}
    X_{ij} = \delta_{ij} p_i + \delta_{i-1,j} e^{q_j - q_i} + \delta_{i+1,j} e^{q_i - q_j},
\end{equation}
which is the Lax matrix for $A_{n - 1}$ non-relativistic open Toda chain. 

The symplectic form on the reduced phase space (\ref{cotangent_bundle_symplectic_form}) is computed directly
\begin{equation}
    \label{symplectic-form-A}
    \omega_{red} = d \mathrm{Tr} (X e^{-q} d(e^q)) = d \mathrm{Tr}(X dq) =
        \mathrm{Tr} (dX \wedge dq) = \sum_{i = 1}^n dp_i \wedge dq_i.
\end{equation}
The Poisson-commuting Hamiltonians on the reduced space can be obtained from Poisson-commuting 
invariant Hamiltonians 
\begin{equation}
    \label{Toda-Hamiltonians-A}
    H_k = \frac{1}{k} \mathrm{Tr} X^k.
\end{equation}
In particular, we have 
\begin{equation}
    H_2 = \sum_{i = 1}^n \frac{p_i^2}{2} + \sum_{i = 1}^{n - 1} e^{2 q_i - 2 q_{i + 1}},
\end{equation}
which is a Hamiltonian of open non-relativistic Toda chain. Let us note that the second family
of invariant Poisson-commuting Hamiltonians become trivial in this gauge
\begin{equation}
    J_k = m_k (g g^{\dagger}) = \prod_{i = 1}^k e^{2q_{n + 1 - i}}.
\end{equation}

As a result, we have the reduced space $M_\lambda = \{(p, q) \in \mathbb{R}^{2n}\}$ with
the canonical symplectic form $\omega_{red}$ (\ref{symplectic-form-A}) and the integrable 
system on $M_\lambda$ determined by Hamiltonians (\ref{Toda-Hamiltonians-A}), which is
nothing but the usual open Toda chain.

\subsection{Moser gauge}
In this section, we study the Hamiltonian reduction from the other point of view. 
Note that (\ref{moment2A}) tells us that matrix $X$ is Hermitian, which means that we can
always diagonalize matrix $X$ by a unitary matrix to bring it to form 
\begin{equation}
    X = \hat{q} = \mathrm{diag}(\hat{q}_1, \ldots, \hat{q}_n),
\end{equation}
where elements $\hat{q}_i$ are real and we assume that they are ordered 
$\hat{q}_1 > \hat{q}_2 > \ldots > \hat{q}_n$, which could always be achieved by the Weyl group action.
The residual group action consists of $N_{+} \times U(1)^n$, 
we will use $N_{+}$ to bring matrix $g$ to lower triangular form via Gauss decomposition, 
at least on the dense subspace of $G$ and $U(1)^{n}$ action makes the diagonal elements of matrix
$g$ real. 

Thus, we are left only with the first momentum map equation (\ref{moment1A}) which can be 
written explicitly as
\begin{equation}
    g \hat{q} g^{-1} = \lambda_{\mathfrak{n}_{-}} + \hat{q},
\end{equation}
which is equivalent to 
\begin{equation}
\label{mom}
    g \hat{q} - \lambda_{\mathfrak{n}_{-}} g - \hat{q} g = 0.
\end{equation}
Then, we have a recursive formula for the elements of $g$
\begin{equation}
    (\hat{q}_j - \hat{q}_i) g_{ij} = g_{i - 1, j},
\end{equation}
which expresses lower-diagonal elements of $g$ via the diagonal ones $g_{ii} = \hat{a}_i$, 
which were chosen to be real,  
\begin{equation}
    g_{ij} = \frac{g_{i - 1, j}}{\hat{q}_j - \hat{q}_i} = 
        \hat{a}_j \cdot \prod_{k = j + 1}^i \frac{1}{\hat{q}_j - \hat{q}_k}, 
            \quad j \leq i.
\end{equation}

In this gauge, nontrivial Poisson-commuting Hamiltonians on the reduced space come from 
invariant Poisson-commuting functions $m_k(g g^{\dagger})$ (\ref{J}). To compute these minors
explicitly we use the Cauchy--Binet formula, which in this case states that
\begin{equation}
    m_k(g g^{\dagger}) = 
        \sum_{i_1 < i_2 < \ldots < i_k } 
            |V_{i_1, \ldots, i_k}(\hat{q}_1, \ldots, \hat{q}_n, \hat{a}_1, \ldots, \hat{a}_n)|^2,
\end{equation}
which is computed thanks to (\ref{detV}) 
\begin{equation}
\label{GoldA}
    \hat{H}_k = m_k(g g^{\dagger}) = 
        \sum_{|I| = k}
            \prod_{l \in I} \hat{a}^2_l
            \prod_{\substack{i \in I, j \notin I \\ j > i}}
                \frac{1}{(\hat{q}_i - \hat{q}_j)^2},
\end{equation}
where the sum goes over ordered sets $I = (i_1, \ldots , i_k)$ of $k$ elements. 
The symplectic form on the reduced space is easily computed 
\begin{equation}
\label{reduced_form}
    \hat{\omega}_{red} = 
        \mathrm{Tr}(d\hat{Q} \wedge g^{-1} dg) - \mathrm{Tr}(\hat{Q} g^{-1} dg \wedge g^{-1} dg) =  
        \sum_{i = 1}^n d \hat{q}_i \wedge \frac{d\hat{a}_i}{\hat{a}_i}.
\end{equation}
To bring Hamiltonians (\ref{GoldA}) to a more recognizable form we change the variables
\begin{equation}
    e^{2 \hat{p}_i} = \hat{a}_i^2 \cdot
        \prod_{j = i + 1}^n (\hat{q}_i - \hat{q}_j) \cdot 
        \prod_{k = 1}^{i - 1} \frac{1}{\hat{q}_k - \hat{q}_i}.
\end{equation}
Thus, in new coordinates, the symplectic form (\ref{reduced_form}) on reduced space is written as
\begin{equation}
    \hat{\omega}_{red} = \sum_{i = 1}^n d\hat{q}_i \wedge d\hat{p}_i
\end{equation}
with Hamiltonians (\ref{GoldA}) in new coordinates being
\begin{equation}
\label{HamiltoniansA}
    \hat{H}_k = 
        \sum_{|I| = k} 
            \prod_{\substack{i \in I \\ j \notin I}}
                \frac{1}{|\hat{q}_i - \hat{q}_j|}
            \prod_{l \in I} e^{2 \hat{p}_l},
\end{equation}
where we use modules for the sake of convenience.
For instance, the first Hamiltonian of the rational Goldfish model is
\begin{equation}
    \hat{H}_1 = \sum_{i = 1}^n e^{2 \hat{p}_i}
        \prod_{\substack{j = 1 \\ j \neq i}}^n \frac{1}{|\hat{q}_i - \hat{q}_j|}.
\end{equation}
Let us also note that the Hamiltonians (\ref{HamiltoniansA}) can be obtained 
of as a strong coupling limit of the rational Ruijsenaars--Schneider system. 
Recall that the rational Ruijsenaars--Schneider Hamiltonians are
\begin{equation*}
    \label{RuijA}
    H_k^{\mathrm{RS}} = 
        \sum_{|I| = k} \prod_{\substack{i \in I \\ j \notin I}} 
            \frac{\hat{q}_i - \hat{q}_j + \nu}{\hat{q}_i - \hat{q}_j} 
                \prod_{l \in I} e^{2 \hat{p}_l}.
\end{equation*}
Then we have the following relation between the Ruijsenaars--Schneider Hamiltonians and
Goldfish Hamiltonians
\begin{equation*}
    \lim_{\nu \to \infty} \frac{H_k^{\mathrm{RS}}}{\nu^{k(n-k)}} = \hat{H}_k.
\end{equation*}

\section{\( C_n \) case}
We start with the $C_n$ root system because technically it is the simplest one
apart from the $A_n$ case. 
In this case we consider $G = Sp(2n, \mathbb{C})$ --- a complex symplectic group
of matrices which preserve the symplectic bilinear form \(\Omega\)
\begin{equation}
\label{form}
    G = Sp(2n, \mathbb{C}) = 
        \{g \in GL(2n, \mathbb{C}) \mid g \Omega g^T = \Omega\}, \quad
    \Omega = \begin{pmatrix}
        0 & P \\
        -P & 0
    \end{pmatrix}, 
\end{equation}
where \(P\) is the \(n \times n\) matrix with elements \(P_{ij} = \delta_{n + 1 - i, j}\).
The corresponding Lie algebra \(\mathfrak{g} = \mathfrak{sp}(2n)\) consists matrices
of the form
\begin{equation}
    \mathfrak{g} = \{X \in \mathfrak{gl}(2n) \mid X \Omega + \Omega X = 0\}.
\end{equation}
Maximal compact subgroup $K = Sp(2n, \mathbb{C}) \cap U(2n)$ and the subgroups \(N_+\)
consists of upper-triangular unipotent symplectic matrices of the form
\begin{equation}
    \label{NC}
    \begin{pmatrix}
        A & B \\
        0 & P (A^T)^{-1} P
    \end{pmatrix},
\end{equation}
where $A$ is a unipotent upper-triangular matrix \(n \times n\) matrix, and
\(B\) is an arbitrary \(n \times n\) matrix. Note that matrix (\ref{NC}) is also 
upper-triangular in the usual sense of \(2n \times 2n\) matrices due to the 
appropriate choice of \(\Omega\). This property is useful for the computations
in the Moser gauge. 

The momentum map equation related to the action of \(N_+\) has the same form as
in type \(A\), but with the sum over the simple roots of \(C_n\) system. In this
matrix representation it can be written explicitly as
\begin{equation}
    \label{momentum_C}
    \mathrm{Pr}_{\mathfrak{n}_-}(g X g^{-1}) = \lambda_{\mathfrak{n}_-} =
        \sum_{\alpha \in \Pi} e_{-\alpha} = 
        \begin{pmatrix}
            I & 0 \\
            I_1 & -I
        \end{pmatrix},
\end{equation}
where \(I\) and \(I_1\) are matrices with the elements \(I_{ij} = \delta_{i - 1, j}\)
and $(I_1)_{ij} = \delta_{i, 1} \delta_{j, n}$. For some details about root systems and
matrix representations of the corresponding algebras, one can refer to Appendix 1.

Let us also note that \(\mathfrak{k} = \mathrm{Lie}(K)\) consist of matrices
\begin{equation}
    \begin{pmatrix}
        d & c \\
        -c^\dagger & P d^T P 
    \end{pmatrix},
\end{equation}
where \(d\) and \(c\) are \(n \times n\) matrices, such that $d^\dagger = -d$ and
\(P c^T P = c\). The momentum map equation related to the action of \(K\) restricts
\(X\) to lie in the orthogonal complement to the Lie subalgebra \(\mathfrak{k}\), then 
every \(X\) from the preimage of the momentum map has the form
\begin{equation}
    \label{hermitianC}
    \mathrm{Pr}_{\mathfrak{k}}(X) = 0 \quad \Longleftrightarrow \quad
    X = \begin{pmatrix}
        a & b \\
        b^\dagger & -P a^T P
    \end{pmatrix},
\end{equation}
where \(a\) and \(b\) are \(n \times n\) matrices satisfying \(a^\dagger = -a\) and
\(b = P b^T P\) respectively. Let us stress that we can always diagonalize a matrix
of the form (\ref{hermitianC}) by the adjoint action of unitary-symplectic group \(K\).

Now, let us turn to the description of the Hamiltonian reduction for the symplectic group.

\subsection{Toda gauge}
Using the Iwasawa decomposition and the action of \(N_+ \times K\) we diagonalize the 
matrix $g$ to bring it to the form 
\begin{equation}
    g = e^q = e^{\sum_{i = 1}^n q_i h_i} = 
    \begin{pmatrix}
        e^Q & 0 \\
        0 & P e^{-Q} P
    \end{pmatrix} =
    \mathrm{diag}(e^{q_1}, \ldots, e^{q_n}, e^{-q_n}, \ldots, e^{-q_1}),
\end{equation}
where $q_i$ are real. Since we know that matrix $X$ is of the form (\ref{hermitianC}) 
we can rewrite the first momentum map equation (\ref{momentum_C}) explicitly in terms
of the matrix elements of \(a\) and \(b\)
\begin{gather}
    e^{q_i - q_j} a_{ij} = \delta_{i - 1, j}, \quad i < j, \\
    e^{-q_{n + 1 - i} - q_j} b^\dagger_{ij} = \delta_{i, 1} \delta_{j, n},
\end{gather}
which enables us to find matrix $X$ uniquely up to its diagonal elements. After the
computations we get that matrix $X$ is exactly the Lax matrix of $C_n$ non-relativistic 
open Toda chain
\begin{equation}
    a_{ij} = \delta_{ij} p_i + \delta_{i - 1, j} e^{q_{i - 1} - q_i} + 
        \delta_{j - 1, i} e^{q_{j - 1} - q_j}, \quad
    b_{ij} = \delta_{j, 1} \delta_{i, n} e^{2q_n},
\end{equation}
or, in Lie algebraic terms, \(X\) has the same form as in the case of \(A_{n - 1}\).
\begin{equation}
    X = \sum_{i = 1}^n p_i h_i + \sum_{\alpha \in \Pi} e^{\alpha(q)} (e_\alpha + e_{-\alpha}).
\end{equation}

The symplectic form on the reduced space is easily computed
\begin{equation}
    \omega_{red} = \mathrm{Tr} (dX \wedge g^{-1}dg) = 
        2 \sum_{i = 1}^n dp_i \wedge dq_i,
\end{equation}
with the first non-trivial Hamiltonian being
\begin{equation}
    H_2 = \frac{1}{4} \mathrm{Tr}(X^2) = 
        \sum_{i = 1}^n \frac{p_i^2}{2} + 
            \sum_{i = 1}^{n - 1} e^{2q_i - 2q_{i + 1}} + 
                \frac{1}{2} e^{4q_n}.
\end{equation}
The higher Hamiltonians of \(C_n\) Toda chain are given by the traces of the even powers 
of \(X\) 
\begin{equation}
    H_k = \frac{1}{4k} \mathrm{Tr}(X^{2k}), \quad 1 \le k \le n.
\end{equation}

\subsection{Moser gauge}
Let us switch to the alternative description of the reduced space. Firstly, we note that
due to the momentum map equations, we can diagonalize the matrix $X$ by conjugation of the
group \(K\) to bring it to the form
\begin{equation}
    X = \hat{q} = \sum_{i = 1}^n \hat{q}_i h_i = 
    \begin{pmatrix}
        \hat{Q} & 0 \\
        0 & -P \hat{Q} P
    \end{pmatrix} = 
    \mathrm{diag}(\hat{q}_1, \ldots, \hat{q}_n, -\hat{q}_n, \ldots, -\hat{q}_1),
\end{equation}
where we assume that coordinates $\hat{q}_i$ belong to the following Weyl chamber
$\hat{q}_1 > \hat{q}_2 > \ldots > \hat{q}_n > 0$, 
which could be achieved by the action of the Weyl group,
and we use the action of unipotent group \(N_+\) to bring matrix $g$ to the Borel
subgroup \(B_-\)
\begin{equation}
    g = \begin{pmatrix}
        A & 0 \\
        C & P (A^{T})^{-1} P
    \end{pmatrix},
\end{equation}
where matrix $A$ is lower-triangular. Then, the momentum map equation (\ref{momentum_C}) 
can be solved using the recurrent relations obtained in a similar fashion (\ref{mom}) 
as in the $A_n$ case
\begin{gather}
    (\hat{q}_j - \hat{q}_i) A_{ij} = A_{i - 1, j}, \quad i < j \\
    (\hat{q}_{n + 1 - i} + \hat{q}_j) C_{ij} - \delta_{i, 1} A_{n, j} + C_{i - 1, j} = 0.
\end{gather}
These equations enable us to find matrices $A$ and $C$ from the diagonal part of the 
matrix \(A\), which is not fixed by the momentum map conditions. Note that we can always
use the residual diagonal symmetry to make diagonal elements of $A$ real. We obtain
\begin{gather}
    A_{ij} = 
        \hat{a}_j \cdot 
            \prod_{k = j + 1}^i \frac{1}{\hat{q}_j - \hat{q}_k}, \qquad
    (P (A^{T})^{-1} P)_{ij} = 
        (-1)^{i - j} \cdot \frac{1}{\hat{a}_{n + 1 - j}} \cdot 
            \prod_{k = j + 1}^i \frac{1}{\hat{q}_{n + 1 - k} - \hat{q}_{n + 1 - j}}, \\
    C_{ij} =
        (-1)^{i - 1} \hat{a}_j \cdot 
            \prod_{k = j + 1}^n \frac{1}{\hat{q}_j - \hat{q}_k} \cdot
                \prod_{m = 1}^i \frac{1}{\hat{q}_{n + 1 - m} + \hat{q}_j}.
\end{gather}
Invariant Poisson-commuting Hamiltonians are $m_k(g g^{\dagger})$ --- principal  
right-lowest minors of size $k$ for $1 \leq k \leq n$, which can be computed now
using the Cauchy--Binet formula.  We will take the sum over all the minors of size
\(k\) constructed from lowest \(k\) rows and \(s = 0, \ldots, k\) columns with indices
\(I = \{i_1 < \ldots < i_s\}\) from \(1, \ldots, n\) set and \((k - s)\) columns with indices 
\(n + J = \{n + j_1 < \ldots < n + j_{k - s}\}\) from \(n + 1, \ldots, 2n\) set
\begin{equation}
    m_k(g g^\dagger) = \sum_{\substack{I, J \\ |I| + |J| = k}} 
        |V_{i_1, \ldots, i_s, n + j_1, \ldots, n + j_{k - s}}|^2
\end{equation}

We will need only the lowest \(n\) rows of the matrix \(g\) to compute all these
\(C_n\) Goldfish Hamiltonians. Note that these \(n\) lowest rows are the same as 
\(n\) lowest rows of the \(2n \times 2n\) matrix of the type \(M(b, x)\) (\ref{M_matrix}),
with parameters 
\(b = \{(-1)^n \hat{a}_1, \ldots, (-1)^{n - i + 1} \hat{a}_i, \ldots, -\hat{a}_n, 
-\hat{a}_n^{-1}, \ldots, -\hat{a}_1^{-1}\}\) and 
\(x = \{-\hat{q}_1, \ldots, -\hat{q}_n, \hat{q}_n, \ldots, \hat{q}_1\}\)
then due to (\ref{detV}) we obtain for the module of this minor
\(|V_{IJ}| = |V_{i_1, \ldots, i_s, n + j_1, \ldots, n + j_{k - s}}|\)
\begin{equation}
    |V_{IJ}| = 
        \left|
            \frac{
                \prod\limits_{l = 1}^s \hat{a}_{i_l} \cdot
                \prod\limits_{r = 1}^{k - s} \hat{a}_{n + 1 - j_r}^{-1}
            }{
                \prod\limits_{l = 1}^s \Big(
                    \prod\limits_{\substack{m = i_l + 1 \\ m \notin I}}^n
                        (\hat{q}_{i_l} - \hat{q}_m) \cdot
                    \prod\limits_{\substack{p = 1 \\ p \notin J}}^n
                        (\hat{q}_{i_l} + \hat{q}_{n + 1 - p})
                \Big) \cdot
                \prod\limits_{r = 1}^{k - s} 
                    \prod\limits_{\substack{p = j_r + 1 \\ p \notin J}}^n
                        (\hat{q}_{n + 1 - j_r} - \hat{q}_{n + 1 - p})
            }
        \right|
\end{equation}
Now to compute the Hamiltonians we sum over all tuples $I = (i_1 < \ldots < i_k)$ and
$J = (j_1 < \ldots < j_m)$
\begin{equation}
    \label{hamiltonianC}
    \hat{H}_k = \sum_{\substack{I, J \\ |I| + |J| = k}} |V_{IJ}|^2.
\end{equation}
To bring Hamiltonians (\ref{hamiltonianC}) into the recognizable form we make a change of 
variables
\begin{equation}
    \hat{a}_i^2 = e^{2 \hat{p}_i} \cdot
        \prod_{j = i + 1}^n (\hat{q}_i - \hat{q}_j) \cdot
        \prod_{k = 1}^n (\hat{q}_i + \hat{q}_k) \cdot
        \prod_{l = 1}^{i - 1} \frac{1}{\hat{q}_l - \hat{q}_i}.
\end{equation}
After simplifications, we obtain the Hamiltonians
\begin{multline}
    \hat{H}_k = \sum_{\substack{I, J \\ |I| + |J| = k}}
        \prod\limits_{l = 1}^s 
        \frac{
            e^{2 \hat{p}_{i_l}}  
        }{
            \prod\limits_{\substack{m = 1 \\ m \notin I}}^n 
                |\hat{q}_{i_l} - \hat{q}_m| 
            \prod\limits_{\substack{p = 1 \\ p \notin J}}^n 
                |\hat{q}_{i_l} + \hat{q}_{n + 1 - p}|
        } \times \\ \times
        \prod\limits_{r = 1}^{k - s} 
        \frac{
            e^{-2 \hat{p}_{n + 1 - j_r}}
        }{
            \prod\limits_{\substack{m = 1 \\ m \notin I}}^n
                |\hat{q}_{n + 1 - j_r} + \hat{q}_m|           
            \prod\limits_{\substack{p = 1 \\ p \notin J}}^n
                |\hat{q}_{n + 1 - j_r} - \hat{q}_{n + 1 - p}|
        }.
\end{multline}
For example, the first Hamiltonian is
\begin{equation}
    \hat{H}_1 = \sum_{i = 1}^n (e^{2 \hat{p}_i} + e^{-2 \hat{p}_i}) 
    \frac{1}{2 |\hat{q}_i |} 
        \prod_{\substack{j = 1 \\ j \ne i}}^n 
            \frac{1}{|\hat{q}_i - \hat{q}_j| |\hat{q}_i + \hat{q}_j|}.
\end{equation}
The symplectic form after reduction is canonical
\begin{equation}
    \hat{\omega}_{red} = 2\sum\limits_{i=1}^n d\hat{q}_i \wedge d \hat{p}_i.
\end{equation}

\section{$B_n$ case}
In this case we consider a complex orthogonal group $G = SO(2n+1, \mathbb{C})$, which
consists of \((2n + 1) \times (2n + 1)\) matrices preserving the form
\begin{equation}
    \Omega = \begin{pmatrix}
        0_{n, n} & 0_{n, 1} & P \\
        0_{1, n} & 1 & 0_{1, n} \\
        P & 0_{n, 1} & 0_{n, n}
    \end{pmatrix},
\end{equation}
where $P$ is the same as in the (\ref{form}). The maximal compact subgroup is 
$K = SO(2n + 1, \mathbb{C}) \cap U(2n + 1)$ and unipotent group $N_{+}$ is formed by
orthogonal matrices of the form
\begin{equation}
    \begin{pmatrix}
    A & C_1 & B \\
    0 & 1 & C_2 \\
    0 & 0 & P (A^T)^{-1}     
    \end{pmatrix},
\end{equation}
where \(A\) is an \(n \times n\) unipotent upper-triangular matrix, and some constraints
on \(A, B, C_1\) and \(C_2\) comes from the orthogonality of this matrix.
We choose an invariant bilinear form to be a trace form. 

The momentum map equations 
for \(N_+\) action are determined by the value of the momentum map (\ref{moment1A}) 
\begin{equation}
    \lambda = 
    \begin{pmatrix}
    I & 0 & 0 \\
    I_1 & 0 & 0 \\
    0 & I_2 & -I
    \end{pmatrix},
\end{equation}
where $I_{i, j} = \delta_{i - 1, j}$, $(I_1)_{i} = \delta_{i, n}$ and $(I_2)_{j} = -\delta_{j, 1}$.
The momentum map equations for \(K\) action restricts \(X\) to lie in the complement to 
$\mathfrak{k} = \mathrm{Lie}(K)$, i.e. be of the form
\begin{equation}
    X =  
    \begin{pmatrix}
    a & c & b \\
    c^{\dagger} & 0 & -c^T P \\
    b^{\dagger} & -P \bar{c} & -P a^T P
    \end{pmatrix},
\end{equation}
where $b = -P b^T P$ and $a^{\dagger} = a$. Again, we can always diagonalize such a matrix using
the conjugation by \(K\). Now we turn to the computation of the reduction.

\subsection{Toda gauge}
By the Iwasawa decomposition, we diagonalize the matrix $g$, bringing it to the form 
\begin{equation}
    g = \mathrm{diag}(e^{Q}, 1, P e^{-Q} P),
\end{equation}
where $q_i$ are real numbers. Now we can explicitly solve the first momentum map equation.
\begin{equation}
    a_{ij}  e^{q_i - q_j} = \delta_{i - 1, j}, \quad i < j, \qquad
    b_{ij} = 0, \qquad
    c^{\dagger}_i e^{-q_i} = \delta_{i, n}.
\end{equation}
This enables us to calculate the entries of matrix $X$ up to its Cartan factor
\begin{equation}
    a_{ij} = \delta_{ij} p_i + \delta_{i - 1, j} e^{q_{i - 1} - q_i} +
        \delta_{j - 1, i} e^{q_{j - 1} - q_j}, \quad
    c_i = \delta_{i, n} e^{q_n},
\end{equation}
which gives us exactly the Lax matrix for type $B_n$ non-relativistic open Toda chain. 
The reduced symplectic form is 
\begin{equation}
    \omega_{red} = 2 \sum_{i = 1}^n dp_i \wedge dq_i
\end{equation}
with the first non-trivial Hamiltonian being
\begin{equation}
    H_2 = \frac{1}{4} \mathrm{Tr}(X^2) = 
        \sum_{i = 1}^n \frac{p_i^2}{2} + 
            \sum_{i = 1}^{n - 1} e^{2q_i - 2q_{i + 1}} + e^{2q_n}.
\end{equation}
The higher Hamiltonians are as usual
\begin{equation}
    H_{2k} = \frac{1}{4k}\mathrm{Tr}(X^{2k}).
\end{equation}

\subsection{Moser gauge}
Here we follow the same footpath as in the $C_n$ case. By conjugating matrix $X$ using maximal 
compact subgroup \(K\) we bring it to the form
\begin{equation}
    X = \mathrm{diag}(\hat{Q}, 0, -P \hat{Q} P) = 
        \mathrm{diag}(\hat{q}_1, \ldots, \hat{q}_n, 0, -\hat{q}_n, \ldots, -\hat{q}_1),
\end{equation}
where we assume that coordinates $\hat{q}_i$ belong to the following Weyl chamber
$\hat{q}_1 > \hat{q}_2 > \ldots > \hat{q}_n > 0$, 
which could be achieved by the action of the Weyl group.
Then, using Gauss decomposition we bring matrix $g$ to the lower-triangular form
\begin{equation}
    g =
    \begin{pmatrix}
    A & 0 & 0 \\
    C_1 & 1 & 0 \\
    C & C_2 & P (A^T)^{-1} P 
    \end{pmatrix}.
\end{equation}
Now we can solve the momentum map equations to obtain the recurrence relations on matrix elements of
\(A, C, C_1, C_2\), which could be expressed via the diagonal elements of matrix $A$ denoted as
\(\hat{a}_1, \ldots, \hat{a}_n\)
\begin{gather}
    A_{ij} = \hat{a}_j \prod_{k = j + 1}^i \frac{1}{\hat{q_j} - \hat{q}_k}, \quad
    (P (A^T)^{-1} P)_{ij} = (-1)^{i - j} \frac{1}{\hat{a}_{n + 1 - j}} 
        \prod_{k = j + 1}^i \frac{1}{\hat{q}_{n + 1 - k} - \hat{q}_{n + 1 - j}}, \\
    (C_1)_j = \frac{\hat{a}_j}{\hat{q}_j} \prod_{k = j + 1}^n \frac{1}{\hat{q}_j - \hat{q}_k}, \quad
    (C_2)_i = - \prod_{k = 1}^i \frac{(-1)}{\hat{q}_{n + 1 - k}}, \quad
    C_{ij} = (-1)^i \hat{a}_j \prod_{k = j + 1}^n \frac{1}{\hat{q}_j - \hat{q}_k} 
        \prod_{l = 1}^i \frac{1}{\hat{q}_j + \hat{q}_{n + 1 - l}}.
\end{gather}

Invariant Poisson-commuting Hamiltonians are $m_k(gg^{\dagger})$ --- principal right-lowest minors
of size $k$ for $1 \leq k \leq n$, which can be computed using Cauchy--Binet formula.

In this case, we have to compute minors of two different types, ones without the \((n + 1)\)-th column,
and the others with this exceptional column. The minors have again the form (\ref{detV1}), 
and are constructed in the same manner as in the $C_n$ case.

Denote by $V_{IJ} = V_{i_1, \ldots, i_s, n + 1 + j_1, \ldots, n + 1 + j_{k - s}}$ the minor 
constructed from the columns labeled by \(I = \{i_1 < \ldots < i_s\}\) from the set 
\(\{1, \ldots, n\}\) and the columns labeled by 
\(n + 1 + J = \{n + 1 + j_1 < \ldots < n + 1 + j_{k - s}\}\). Applying formula (\ref{detV1}),
we get
\begin{equation}
    |V_{IJ}| = \left| \frac{
        \prod\limits_{l = 1}^s \hat{a}_{i_l} 
        \prod\limits_{r = 1}^{k - s} \hat{a}_{n + 1 - j_r}^{-1}
    }{
        \prod\limits_{l = 1}^s \Big(
            \hat{q}_{i_l}
            \prod\limits_{\substack{m = i_l + 1 \\ m \notin I}}^n 
                (\hat{q}_{i_l} - \hat{q}_m)
            \prod\limits_{\substack{p = 1 \\ p \notin J}}^n 
                (\hat{q}_{i_l} + \hat{q}_{n + 1 - p})
        \Big)
        \prod\limits_{r = 1}^{k - s} 
            \prod\limits_{\substack{p = j_r + 1 \\ p \notin J}}^n
                (\hat{q}_{n + 1 - j_r} - \hat{q}_{n + 1 - p})
    } \right|.
\end{equation}
Denote by \(\hat{V}_{IJ} = V_{i_1 , \ldots, i_s, n + 1 , j_1, \ldots, j_{k - s - 1}}\) the minor
constructed from the columns labeled by \(I = \{i_1 < \ldots < i_s\}\) from the set 
\(\{1, \ldots, n\}\), the column \((n + 1)\) and the columns labeled by 
\(n + 1 + J = \{n + 1 + j_1 < \ldots < n + 1 + j_{k - s - 1}\}\). This minor equals to
\begin{equation}
    |\hat{V}_{IJ}| = \left| \frac{ 
        \prod\limits_{l = 1}^s \hat{a}_{i_l} 
        \prod\limits_{r = 1}^{k - s - 1} \hat{a}_{n + 1 - j_r}^{-1}
    }{
        \prod\limits_{\substack{p = 1 \\ p \notin J}}^n \hat{q}_p
        \prod\limits_{l = 1}^k \Big(
            \prod\limits_{\substack{m = i_l + 1 \\ l \notin I}}^n
                (\hat{q}_{i_l} - \hat{q}_m)
            \prod\limits_{\substack{p = 1 \\ p \notin J}}^n
                (\hat{q}_{i_l} + \hat{q}_{n + 1 - p}) 
        \Big)
        \prod\limits_{r = 1}^{k - s - 1} 
            \prod\limits_{\substack{p = j_r + 1 \\ p \notin J}}^n
                (\hat{q}_{n + 1 - j_r} - \hat{q}_{n + 1 - p})
    } \right|.
\end{equation}
The Hamiltonians are constructed as the sum of squares of all minors of the same size
\begin{equation}
    \hat{H}_k = \sum_{\substack{I, J \\ |I| + |J| = k}} |V_{IJ}|^2 + 
        \sum_{\substack{I, J \\ |I| + |J| = k - 1}} |\hat{V}_{IJ}|^2.
\end{equation}
Before writing the Hamiltonians explicitly we make a change of the variables from \(\hat{a}_i\)
to momenta \(\hat{p}_i\)
\begin{equation}
    \hat{a}_i^2 = e^{2 \hat{p}_i} \hat{q}_i \cdot
        \prod_{j = i + 1}^n (\hat{q}_i - \hat{q}_j) \cdot
        \prod_{k = 1}^n (\hat{q}_i + \hat{q}_k) \cdot
        \prod_{l = 1}^{i - 1} \frac{1}{\hat{q}_l - \hat{q}_i},
\end{equation}
which brings the Hamiltonians to the form
\begin{multline} 
    \hat{H}_k = 
        \sum_{\substack{I, J \\ |I| + |J| = k}} 
            \prod_{l = 1}^s \frac{
                e^{2 \hat{p}_{i_l}} / \hat{q}_{i_l}
            }{
                \prod\limits_{\substack{m = 1 \\ m \notin I}}^n 
                    |\hat{q}_{i_l} - \hat{q}_m| 
                \prod\limits_{\substack{p = 1 \\ p \notin J}}^n 
                    |\hat{q}_{i_l} + \hat{q}_{n + 1 - p}|
            } 
            \prod_{r = 1}^{k - s} \frac{
                e^{-2 \hat{p}_{n + 1 - j_r}} / \hat{q}_{n + 1 - j_r}
            }{
                \prod\limits_{\substack{m = 1 \\ m \notin I}}^n
                    |\hat{q}_{n + 1 - j_r} + \hat{q}_m|
            \prod\limits_{\substack{p = 1 \\ p \notin J}}^{n} 
                |\hat{q}_{n + 1 - j_r} - \hat{q}_{n + 1 - p}|
            }
    + \\ + 
    \sum_{\substack{I, J \\ |I| + |J| = k - 1}} 
            \frac{1}{
                \prod\limits_{\substack{m = 1 \\ m \notin I}}^n |\hat{q}_m|
                \prod\limits_{\substack{p = 1 \\ p \notin J}}^n |\hat{q}_{n + 1 - p}|
            }
            \prod_{l = 1}^s \frac{
                e^{2 \hat{p}_{i_l}} 
            }{
                \prod\limits_{\substack{m = 1 \\ m \notin I}}^n 
                    |\hat{q}_{i_l} - \hat{q}_m| 
                \prod\limits_{\substack{p = 1 \\ p \notin J}}^n 
                    |\hat{q}_{i_l} + \hat{q}_{n + 1 - p}|
            } \times \\ \times 
            \prod_{r = 1}^{k - s - 1} \frac{
                e^{-2 \hat{p}_{n + 1 - j_r}} 
            }{
            \prod\limits_{\substack{m = 1 \\ m \notin I}}^n
                |\hat{q}_{n + 1 - j_r} + \hat{q}_m|
                \prod\limits_{\substack{p = 1 \\ p \notin J}}^n
                |\hat{q}_{n + 1 - j_r} - \hat{q}_{n + 1 - p}|
            } 
\end{multline}
For example, the first Hamiltonian is
\begin{equation}
    \tilde{H}_1 = \sum_{i = 1}^n (e^{2 \hat{p}_i} + e^{-2 \hat{p}_i}) 
        \frac{1}{2 \hat{q}_i^2} 
            \prod_{\substack{j = 1 \\ j \neq i}}^n 
                \frac{1}{|\hat{q}_i - \hat{q}_j| |\hat{q}_i + \hat{q}_j|} + 
        \prod_{i = 1}^n \frac{1}{\hat{q}_i^2}.
\end{equation}
The symplectic form on the reduced space in coordinates \(\{\hat{p}, \hat{q}\}\) is the canonical one
\begin{equation}
    \tilde{\omega}_{red} = 2 \sum_{i = 1}^n d \hat{q}_i \wedge d \hat{p}_i.
\end{equation}

\section{$D_n$ case}
In this case we consider $G = SO(2n , \mathbb{C})$ --- a complex orthogonal group preserving symmetric form
\begin{equation}
    \Omega = 
    \begin{pmatrix}
    0 & P \\
    P & 0
    \end{pmatrix},
\end{equation}
where $P_{ij} = \delta_{n + 1 - i, j}$ as before. The maximal compact subgroup is 
$K = SO(2n ,\mathbb{C}) \cap U(2n)$ and the subgroup $N_{+}$ consists of orthogonal matrices of the form
\begin{equation}
    \begin{pmatrix}
    A & B \\
    0 & P (A^T)^{-1} P
    \end{pmatrix},
\end{equation}
with the upper-triangular \(A\). The invariant bilinear form is chosen to trace form. 

The value of the momentum map for \(N_+\)-action is chosen to be
\begin{equation}
    \lambda = 
    \begin{pmatrix}
    I & 0 \\
    I_1 & -I
    \end{pmatrix},
\end{equation}
where $I_{ij} = \delta_{i - 1, j}$ and 
$(I_1)_{ij} = \delta_{2, i} \delta_{j, n} - \delta_{i, 1} \delta_{j, n - 1}$. 
The momentum map for \(K\)-action restricts \(X\) to be
\begin{equation}
    X =
    \begin{pmatrix}
    a & b \\
    b^{\dagger} & -P a^T P
    \end{pmatrix},
\end{equation}
where $a^{\dagger} = a$ and $b = - P b^T P$.

\subsection{Toda gauge}
Let us diagonalize the matrix $g$ by the Iwasawa decomposition to the form
\begin{equation}
    g = \mathrm{diag}(e^Q , P e^{-Q} P) = 
        \mathrm{diag}(e^{q_1}, \ldots, e^{q_n}, e^{-q_n}, \ldots, e^{-q_1}),
\end{equation}
where $q_i$ are real numbers. The momentum map equations are written explicitly 
\begin{gather}
    e^{q_i - q_j} a_{ij} = \delta_{i - 1, j}, \quad i < j, \\
    e^{-q_{n + 1 - i} - q_j} b^{\dagger}_{ij} = 
        \delta_{2, i} \delta_{j, n} - \delta_{i, 1} \delta_{j, n - 1},
\end{gather}
which gives us
\begin{gather}
    a_{ij} = \delta_{ij} p_i + 
        \delta_{i - 1, j} e^{q_{i - 1} - q_i} + 
            \delta_{j - 1,i} e^{q_{j - 1} - q_j}, \\
    b_{ij} = e^{q_{n - 1} + q_n} 
        (\delta_{2, i} \delta_{j, n} - 
            \delta_{i, 1} \delta_{j, n - 1}),
\end{gather}
which gives us the Lax matrix of $D_n$ non-relativistic open Toda chain. 

The symplectic form on the reduced space is computed in the same way as in the previous sections
\begin{equation}
    \omega_{red} = 2 \sum_{i = 1}^n dp_i \wedge dq_i,
\end{equation}

The first nontrivial Hamiltonian of \(D_n\) Toda chain is
\begin{equation}
    H_2 = \frac{1}{4} \mathrm{Tr}(X^2) = \sum\limits_{i=1}^n \frac{p_i^2}{2} + \sum\limits_{i=1}^{n-1} e^{q_{i} - q_{i+1}} + e^{q_{n-1} + q_n},
\end{equation}
and higher Hamiltonians in involution with this one are given by
\begin{equation}
   H_{2k} = \frac{1}{4k}\mathrm{Tr}(X^{2k}).
\end{equation}

\subsection{Moser gauge}
Now we diagonalize matrix $X$ by action of maximal compact subgroup bringing it to the form 
\begin{equation}
    X = \mathrm{diag} (\hat{Q} , -P \hat{Q} P) = 
        \mathrm{diag}(\hat{q}_1, \ldots , \hat{q}_n , -\hat{q}_n , \ldots, -\hat{q}_1),
\end{equation}
where $\hat{q}_i$ are real and belong to the positive Weyl chamber 
\begin{equation}
    \{\hat{q}_i - \hat{q}_j > 0, \quad \hat{q}_i + \hat{q}_j > 0 \mid 1 \le i < j \le n\}.
\end{equation}
We use the action of upper-triangular unipotent matrices to bring $g$ to the form
\begin{equation}
     g = 
    \begin{pmatrix}
    A & 0 \\
    C & P (A^T)^{-1} P
    \end{pmatrix},
\end{equation}
where matrix $A$ is lower-triangular. The momentum map equation gives the recurrences
\begin{equation} 
    A_{ij} (\hat{q}_j - \hat{q}_i) = A_{i - 1, j}, \quad
    C_{ij} (\hat{q}_j + \hat{q}_{n + 1 - i}) =
        -C_{i - 1, j} + \delta_{2, i} A_{n, j} - \delta_{i, 1} A_{n - 1, j}, 
\end{equation}
which allows us to express \(g\) in terms of coordinates \(\hat{q}_j\) and diagonal elements of \(A\),
denoted by \(\hat{a}_j\)
\begin{gather}
    A_{ij} = \hat{a}_j 
        \prod_{k = j + 1}^i \frac{1}{\hat{q}_j - \hat{q}_k}, \\
    C_{1j} = -\frac{A_{n - 1, j}}{\hat{q}_j + \hat{q}_n}, \qquad
    C_{ij} = (-1)^{i - 2} \cdot 2\hat{q}_j \cdot A_{nj} 
        \prod_{k = 1}^i \frac{1}{\hat{q}_j + \hat{q}_{n + 1 - k}}.
\end{gather}
Notice that since \(A_{n - 1, n} = 0\), we have $C_{1, n} = 0$, which makes this case significantly
different from the other cases. Because of this observation, we have two different situations when computing
the Hamiltonians $m_k(g g^{\dagger})$ for $1 \leq k \leq n$. The first case is when $k \neq n$, 
in this case, the computations follow from the $C_n$ case with slight modifications. In this case, we get 
the following expression for the minors \(V_{IJ} = V_{i_1, \ldots, i_s, j_1, \ldots, j_{k - s}}\)
constructed from columns \(I = (i_1 < \ldots < i_s)\) from the set \(\{1, \ldots, n\}\) and 
\(n + J = (n + j_1, \ldots, n + j_{k - s})\) from the set \(\{n + 1, \ldots, 2n\}\)
\begin{equation}
    |V_{IJ}| = \left| \frac{
        \prod\limits_{l = 1}^s 
            2 \hat{a}_{i_l} \hat{q}_{i_l} \cdot
        \prod\limits_{r = 1}^{k - s}
            \hat{a}_{n + 1 - j_r}^{-1}
    }{
        \prod\limits_{l = 1}^s \Big(
            \prod\limits_{\substack{m = i_l + 1 \\ m \notin I}}^n
                (\hat{q}_{i_r} - \hat{q}_m)
            \prod\limits_{\substack{p = 1 \\ p \notin J}}^n
                (\hat{q}_{i_r} + \hat{q}_{n + 1 - p})
        \Big)
        \prod\limits_{r = 1}^{k - s} 
            \prod\limits_{\substack{p = j_r + 1 \\ p \notin J}}^n
                (\hat{q}_{n + 1 - j_r} - \hat{q}_{n + 1 - p})
    } \right|.
\end{equation}
for $k < n$. The largest minors $V_{IJ} = |V_{i_1, \ldots, i_s, j_1, \ldots, j_{n - s}}|$
can be computed directly using expansion with respect to the first row, the computations are
straightforward but quite tedious, thus we state the answer
\begin{equation}
    |V_{IJ}| = \left| \frac{
        \prod\limits_{l = 1}^s 
            \hat{a}_{i_l} 
        \prod\limits_{r = 1}^{n - s} 
            \hat{a}_{n + 1 - j_r}^{-1} \cdot
        2^{s - 1} \Big(
            \prod\limits_{l = 1}^s 
                \hat{q}_{i_l} + 
            (-1)^s \prod\limits_{\substack{p = 1 \\ p \notin J}}^n 
                \hat{q}_{n + 1 - p}
        \Big)
    }{
        \prod\limits_{l = 1}^s \Big(
            \prod\limits_{\substack{m = i_l + 1 \\ m \notin I}}^n 
                (\hat{q}_{i_r} - \hat{q}_m) 
            \prod\limits_{\substack{p = 1 \\ p \notin J}}^n
                (\hat{q}_{i_r} + \hat{q}_{n + 1 - p})
        \Big)
        \prod\limits_{r = 1}^{n - s}
            \prod\limits_{\substack{p = j_r + 1 \\ p \notin J}}
                (\hat{q}_{n + 1 - j_r} - \hat{q}_{n + 1 - p})
    } \right|.
\end{equation}
If we sum up the squares of minors we obtain the explicit expressions for the Hamiltonians, 
but before writing them down we again make an almost canonical change of coordinates
to momenta \(\hat{p}_i\)
\begin{equation}
    \hat{a}_i^2 = e^{2 \hat{p}_i} \cdot 
        \prod_{j = i + 1}^n (\hat{q}_i - \hat{q}_j) \cdot
        \prod_{\substack{k = 1 \\ k \ne i}}^n (\hat{q}_i + \hat{q}_k) \cdot
        \prod_{l = 1}^{i - 1} \frac{1}{\hat{q}_l - \hat{q}_i},
\end{equation}
which brings the Hamiltonians to the form
\begin{multline}
    \hat{H}_k = 2^k \sum_{\substack{I, J \\ |I| + |J| = k}} 
        \prod_{l = 1}^s \frac{
            e^{2 \hat{p}_{i_l}} |\hat{q}_{i_l}|
            }{
                \prod\limits_{\substack{m = 1 \\ m \notin I}}^n
                    |\hat{q}_{i_l} - \hat{q}_m|
                \prod\limits_{\substack{p = 1 \\ p \notin J}}^n
                    |\hat{q}_{i_l} + \hat{q}_{n + 1 - p}|
            } \times \\ \times
        \prod_{r = 1}^{k - s} \frac{
            e^{-2 \hat{p}_{n + 1 - j_r}} |\hat{q}_{n + 1 - j_r}|
            }{
                \prod\limits_{\substack{m = 1 \\ m \notin I}}^n
                    |\hat{q}_{n + 1 - j_r} - \hat{q}_m|
                \prod\limits_{\substack{p = 1 \\ p \notin J}}^n
                    |\hat{q}_{n + 1 - j_r} + \hat{q}_{n + 1 - p}|
            }
\end{multline}
for $s < n$. The last Hamiltonian is a more complicated one
\begin{multline}
    \hat{H}_n = 2^{n - 2} \sum_{\substack{I, J \\ |I| + |J| = n}} 
        \prod_{l = 1}^s \frac{
            e^{2 \hat{p}_{i_l}}
            }{
                \prod\limits_{\substack{m = 1 \\ m \notin I}}^n
                    |\hat{q}_{i_l} - \hat{q}_m|
                \prod\limits_{\substack{p = 1 \\ p \notin J}}^n
                    |\hat{q}_{i_l} + \hat{q}_{n + 1 - p}|
            } \times \\ \times
        \prod_{r = 1}^{n - s} \frac{
            e^{-2 \hat{p}_{n + 1 - j_r}}
            }{
                \prod\limits_{\substack{m = 1 \\ m \notin I}}^n
                    |\hat{q}_{n + 1 - j_r} - \hat{q}_m|
                \prod\limits_{\substack{p = 1 \\ p \notin J}}^n
                    |\hat{q}_{n + 1 - j_r} + \hat{q}_{n + 1 - p}|
            }
        \Big(
            \prod_{l = 1}^s \hat{q}_{i_l} +
                (-1)^s \prod_{\substack{p = 1 \\ p \notin J}}^n \hat{q}_{n + 1 - p}
        \Big)^2.
\end{multline}
For example, the first nontrivial Hamiltonian of \(D_n\) Goldfish model is
\begin{equation}
    \hat{H}_1 = \sum_{i = 1}^n 
        (e^{2 \hat{p}_i} + e^{-2 \hat{p}_i})
        \sum_{\substack{j = 1 \\ j \ne i}}^n 
            \frac{1}{|\hat{q}_i - \hat{q}_j| |\hat{q}_i + \hat{q}_j|}.
\end{equation}

The symplectic form on the reduced space is canonical
\begin{equation}
    \hat{\omega}_{red} = 2 \sum_{i = 1}^n d \hat{q}_i \wedge d\hat{p}_i.
\end{equation}

\section{Discussion}
In this paper, we have considered the Hamiltonian reduction of cotangent bundles to
classical complex simple Lie groups of \(B, C, D\) type with respect to Hamiltonian actions of 
unipotent and maximal compact subgroups. Taking two families of invariant 
Poisson-commuting functions of the initial space we arrived at two Ruijsenaars dual
integrable systems on the reduced space: the first systems are non-relativistic Toda chains
related to \(B, C\), and \(D\) root systems, and the second one can be identified with 
\(B, C, D\) rational Goldfish models, which could be viewed as a strong coupling
limit of rational Ruijsenaars--Schneider models.

The first and obvious further question would be to interpret quantum Ruijsenaars duality
between quantum open non-relativistic Toda chains and quantum Goldfish models, which are
degenerations of (rational) Macdonald operators, as a relation coming from quantum 
Hamiltonian reduction. In this case, the duality must be a well-known bispectrality 
between Whittaker functions, which are eigenfunctions for quantum Toda Hamiltonians, 
and eigenfunctions of quantum rational Goldfish Hamiltonians. 

The second open problem is to promote the classical Ruijsenaars duality to the case of
the relativistic open Toda chain, the quasi-Hamiltonian reduction \cite{AMM} and Poisson
reduction descriptions \cite{Lu} of the open relativistic Toda chain were obtained in 
\cite{FM, FT}, but the dual Goldfish system was not derived yet using the reduction procedure,
let us not for some cases the dual Goldfish system was already present in the original work
of Ruijsenaars \cite{R}. 

Another possibly interesting question is to consider the case of exceptional Lie groups
for which Toda systems are well-known for all root systems. Still, the dual systems in the
exceptional case remain unknown to the best of our knowledge. The most complicated question
is to find the dual system for periodic Toda systems. The reduction picture for the periodic
Toda chain is well-known \cite{O}, but the dual system remains elusive, and it is known only
for the two-particle case \cite{ZLB}, which involves the elliptic functions with dynamical modulus. 
Lastly, we anticipate the existence of an interesting Ruijsenaars dual for
full symmetric Toda chain.

\section{Acknowledgments}
We are grateful to M. Feigin, A. Liashyk, Ie. Makedonskyi, N. Reshetikhin, B. Vlaar 
for the valuable comments and discussions. 
M. V. wishes to thank the INI and LMS for the financial support and the School of Mathematics 
at the University of Birmingham for their hospitality, especially Prof. Marta Mazzocco for her
support and guidance during the author's stay at Birmingham. M. V. is also grateful to 
the University of Glasgow for the financial support.

\newpage

\section*{Appendix 1: Root systems and standard representations of classical algebras}
We denote $R$ --- root system, $R_{+}$ --- positive roots, $\Pi$ --- simple roots. 
Let \(e_1, \ldots, e_n\) be an orthonormal basis in the \(n\)-dimensional Euclidean space.
We denote by $E_{ij}$ the matrix with $1$ in $i$-th row and $j$-th column 
and with zeroes elsewhere.

\begin{enumerate}
\item \(A_{n - 1}\) root system.
\begin{gather}
    R = \{e_{i} - e_{j} \mid 1 \leq i \neq j \leq n \}, \\
    R_+ = \{e_{i} - e_{j} \mid 1 \leq i < j \leq n \}, \\
    \Pi = \{e_{i} - e_{i + 1} \mid 1 \leq i < n \},
\end{gather}
The corresponding Lie algebra is $\mathfrak{sl}_n$ --- the Lie algebra of traceless 
\(n \times n\) matrices. The Cartan subalgebra is generated by $n - 1$ traceless diagonal
matrices $h_i = E_{ii} - E_{i + 1, i + 1}$, $1 \le i \le n - 1$. 

One-dimensional root space corresponding to the root $e_i - e_j$ is spanned by $E_{ij}$.

\item \(B_n\) root system.
\begin{gather}
    R = \{e_{i} - e_{j} \mid 1 \leq i \neq j \leq n \} \cup 
        \{\pm (e_{i} + e_{j}) \mid 1 \leq i < j \leq n \} \cup 
        \{\pm e_i \mid 1 \leq i \leq n \}, \\
    R_+ = \{e_{i} - e_{j} \mid 1 \leq i < j \leq n \} \cup 
        \{ e_{i} + e_{j} \mid 1 \leq i < j \leq n \} \cup 
        \{ e_i \mid 1 \leq i \leq n \}, \\
    \Pi = \{e_{i} - e_{i + 1} \mid 1 \leq i < n \} \cup \{e_n\}.
\end{gather}
The corresponding Lie algebra is \(\mathfrak{so}_{2n + 1}\) --- the Lie algebra of 
\((2n + 1) \times (2n + 1)\) matrices \(X\), such that \(X \Omega + \Omega X^T = 0\), 
where
\begin{equation}
    \Omega = \sum_{i = 1}^{2n + 1} E_{2n + 2 - i, i}.
\end{equation}
The Cartan subalgebra is generated by \(n\) matrices  
\(h_i = E_{ii} - E_{2n + 1 - i, 2n + 1 - i}\). 

One-dimensional root spaces are represented by
\begin{align}
    e_i - e_j \colon& \quad E_{ij} - E_{2n + 2 - j, 2n + 2 - i}, \\
    e_i + e_j \colon& \quad E_{i, 2n + 2 - j} - E_{j, 2n + 2 - i}, \\
    -e_i - e_j \colon& \quad E_{2n + 2 - i, j} - E_{2n + 2 - j, i}, \\
    e_i \colon& \quad E_{i, n + 1} + E_{n + 1, 2n + 2 - i}, \\
    -e_i \colon& \quad E_{n + 1, i} + E_{2n + 2 - i, n + 1}. 
\end{align}

\item \( C_n\) root system.
\begin{gather*} 
    R = \{e_i - e_j \mid 1 \leq i \neq j \leq n\} \cup 
        \{\pm (e_i + e_j) \mid 1 \leq i < j \leq n\} \cup 
        \{\pm 2e_i \mid 1 \leq i \leq n\}, \\
    R_+ = \{e_i - e_j \mid 1 \leq i < j \leq n\} \cup
        \{e_i + e_j \mid 1 \leq i < j \leq n \} \cup 
        \{2e_i \mid 1 \leq i \leq n\}, \\
    \Pi = \{e_i - e_{i + 1} | 1 \leq i < n \} \cup \{2e_n\}.
\end{gather*}
The corresponding Lie algebra is \(\mathfrak{sp}_{2n}\) --- the Lie algebra of 
\(2n \times 2n\) matrices \(X\), such that \(X \Omega + \Omega X^T = 0\), 
where
\begin{equation}
    \Omega = \sum_{i = 1}^{n} ( E_{i, 2n + 1 - i} - E_{2n + 1 - i, i}).
\end{equation}
The Cartan subalgebra is generated by \(n\) matrices  
\(h_i = E_{ii} - E_{2n + 1 - i, 2n + 1 - i}\). 

One-dimensional root spaces are represented by
\begin{align}
    e_i - e_j \colon& \quad E_{ij} - E_{2n + 1 - j, 2n + 1 - i}, \\
    e_i + e_j \colon& \quad E_{i, 2n + 1 - j} - E_{j, 2n + 1 - i}, \\
    -e_i - e_j \colon& \quad E_{2n + 1 - i, j} - E_{2n + 1 - j, i}, \\
    2e_i \colon& \quad E_{i, 2n + 1 - i}, \\
    -2e_i \colon& \quad E_{2n + 1 - i, i}.
\end{align}

\item \(D_n\) root system.
\begin{gather}
    R = \{e_i - e_j \mid 1 \leq i \neq j \leq n\} \cup 
        \{\pm (e_i + e_j) \mid 1 \leq i < j \leq n\}, \\
    R_+ = \{e_i - e_j \mid 1 \leq i < j \leq n\} \cup 
        \{e_i + e_j \mid \leq i < j \leq n\}, \\
    \Pi = \{e_i - e_{i + 1} \mid 1 \leq i < n\} 
        \cup \{e_{n-1} + e_n \}.
\end{gather}
The corresponding Lie algebra is \(\mathfrak{so}_{2n}\) --- the Lie algebra of 
\(2n \times 2n\) matrices \(X\), such that \(X \Omega + \Omega X^T = 0\), 
where
\begin{equation}
    \Omega = \sum_{i = 1}^{2n}  E_{i, 2n + 1 - i}.
\end{equation}
The Cartan subalgebra is generated by \(n\) matrices  
\(h_i = E_{ii} - E_{2n + 1 - i, 2n + 1 - i}\). 

One-dimensional root spaces are represented by
\begin{align}
    e_i - e_j \colon& \quad E_{ij} - E_{2n + 1 - j, 2n + 1 - i}, \\
    e_i + e_j \colon& \quad E_{i, 2n + 1 - j} - E_{j, 2n + 1 - i}, \\
    -e_i - e_j \colon& \quad E_{2n + 1 - i, j} - E_{2n + 1 - j, i}.
\end{align}

\end{enumerate}

\section*{Appendix 2: Simple examples}
\subsection*{\( G = GL(2, \mathbb{C})\)}
In Toda gauge, the matrix \(g\) can be diagonalized via 
\(N_+ \times K\) action to
\begin{equation}
    g = \begin{pmatrix}
        e^{q_1} & 0 \\
        0 & e^{q_2}
    \end{pmatrix},
\end{equation}
where \(q_1\) and \(q_2\) are real.
In this gauge, the off-diagonal part of the matrix \(X\) can be found from the momentum 
map equations (\ref{momentum})--(\ref{value}), whilst the diagonal part remains unconstrained
\begin{equation}
    X = \begin{pmatrix}
        p_1 &  e^{q_1 - q_2}\\
        e^{q_1 - q_2} & p_2 
    \end{pmatrix}.
\end{equation}
The only nontrivial Hamiltonian is
\begin{equation}
    H = \frac{1}{2}\mathrm{Tr}(X^2) = \frac{p_1^2 + p_2^2}{2} + e^{2q_1 - 2q_2}, 
\end{equation}
and \(\mathrm{Tr}(X)\) is just a total momentum \(p_1 + p_2\).

In Moser gauge, the matrix \(X\) becomes diagonal under \(K\) action
\begin{equation}
    X = \begin{pmatrix}
        \hat{q}_1 & 0 \\
        0 & \hat{q}_2
    \end{pmatrix},
\end{equation}
while \(g\) can be transformed to the lower-diagonal matrix using remaining \(N_+\) 
action
\begin{equation}
    g = \begin{pmatrix}
        e^{\hat{p}_1} & 0 \\
        \frac{e^{\hat{p}_1}}{\hat{q}_1 - \hat{q}_2} & e^{\hat{p}_2}
    \end{pmatrix}.
\end{equation}
The Hamiltonians of the Goldfish model are
\begin{gather}
    \hat{H}_1 = m_1(g g^{\dagger}) = 
        \frac{e^{2 \hat{p}_1}}{(\hat{q}_1 - \hat{q}_2)^2} + e^{2\hat{p}_2}, \quad
    \hat{H}_2 = m_2 (g g^{\dagger}) = e^{2\hat{p}_1 + 2 \hat{p}_2}
\end{gather}
to make the Hamiltonian more symmetric we can make a canonical change of coordinates
\begin{equation}
    e^{2 \hat{p}_1} \mapsto e^{2 \hat{p}_1} \cdot (\hat{q}_1 - \hat{q}_2), \quad
    e^{2 \hat{p}_2} \mapsto e^{2 \hat{p}_2} \cdot \frac{1}{\hat{q}_2 - \hat{q}_1}, \quad 
    \hat{q}_{1, 2} \mapsto \hat{q}_{1, 2},
\end{equation}
which results in
\begin{equation}
    \hat{H_1} = \frac{e^{2\hat{p}_1}}{\hat{q}_1 - \hat{q}_2} + \frac{e^{2\hat{p}_2}}{\hat{q}_2 - \hat{q}_1}, \quad 
    \hat{H_2} = e^{2\hat{p}_1 + 2 \hat{p}_2}.
\end{equation}

\subsection*{\(G = Sp(4, \mathbb{C})\)} 
In Toda gauge, the matrix \(g\) can be diagonalized via 
\(N_{+} \times K\) action to
\begin{equation}
    g = \begin{pmatrix}
        e^{q_1} & 0 & 0 & 0 \\
        0 & e^{q_2} & 0 & 0 \\
        0 & 0 & e^{-q_2} & 0 \\
        0 & 0 & 0 & e^{-q_1}
    \end{pmatrix},
\end{equation}
where $q_1$ and $q_2$ are real. In this gauge, the matrix \(X\) is given by
\begin{equation}
    X = \begin{pmatrix}
        p_1 & e^{q_1 - q_2} & 0 & 0 \\
        e^{q_1 - q_2} & p_2 & e^{2q_2} & 0 \\
        0 & e^{2 q_2} & -p_2 & -e^{q_1 - q_2} \\
        0 & 0 & -e^{q_1 - q_2} & -p_1
    \end{pmatrix}.
\end{equation}
The first nontrivial Hamiltonian of \(C_2\) Toda chain is
\begin{equation}
    H = \frac{1}{4} \mathrm{Tr}(X^2) = \frac{p_1^2 + p_2^2}{2} +
        e^{2q_1 - 2q_2} + \frac{1}{2} e^{4q_2}.
\end{equation}

In Moser gauge, the matrix $X$ becomes diagonal under $K$ action
\begin{equation}
    X = \begin{pmatrix}
        \hat{q}_1 & 0 & 0 & 0\\
        0 & \hat{q}_2 & 0 & 0 \\
        0 & 0 & -\hat{q}_2 & 0 \\
        0 & 0 & 0 & -\hat{q}_1
    \end{pmatrix},
\end{equation}
while $g$ can be transformed to the lower-diagonal matrix using the remaining \( N_{+}\) action 
\begin{equation}
    g = \begin{pmatrix}
        e^{\hat{p}_1} & 0 & 0 & 0 \\
        \frac{e^{\hat{p}_1}}{\hat{q}_1 - \hat{q}_2} & e^{\hat{p}_2} & 0 & 0 \\
        \frac{e^{\hat{p}_1}}{(\hat{q}^2_1 - \hat{q}^2_2)} & 
            \frac{e^{\hat{p}_2}}{2\hat{q}_2} & e^{-\hat{p}_2} & 0 \\
        -\frac{e^{\hat{p}_1}}{2 \hat{q}_1(\hat{q}^2_1 - \hat{q}^2_2)} &
            -\frac{e^{\hat{p}_2}}{2 \hat{q}_2 (\hat{q}_1 + \hat{q}_2)} &
                - \frac{e^{-\hat{p}_2}}{\hat{q}_1 - \hat{q}_2} & e^{-\hat{p}_1}
    \end{pmatrix}.
\end{equation}
The first Hamiltonian of \(C_2\) Goldfish model is
\begin{gather}
    \hat{H} = m_1(g g^\dagger) = 
        \frac{e^{2 \hat{p}_1}}{4\hat{q}^2_1 (\hat{q}^2_1 - \hat{q}^2_2)^2} + 
            \frac{e^{2 \hat{p}_2}}{4 \hat{q}^2_2 (\hat{q}_1 + \hat{q}_2)^2} + 
                \frac{e^{-2 \hat{p}_2}}{(\hat{q}_1 - \hat{q}_2)^2} + 
                    e^{-2 \hat{p}_1},
\end{gather}
after the canonical transformation
\begin{equation}
    e^{2\hat{p}_1} \mapsto 
        e^{2 \hat{p}_1} 2 \hat{q}_1 (\hat{q}^2_1 - \hat{q}^2_2), \quad 
    e^{2\hat{p}_2} \mapsto 
        e^{2 \hat{p}_2} \frac{2 \hat{q}_2 (\hat{q}_1 + \hat{q}_2)}{\hat{q}_2 - \hat{q}_1}, \quad
    \hat{q}_{1, 2} \mapsto \hat{q}_{1, 2}
\end{equation}
it has the form
\begin{equation}
    \hat{H}_1 = 
        \frac{\cosh 2\hat{p}_1}{\hat{q}_1 (\hat{q}_1 - \hat{q}_2) (\hat{q}_1 + \hat{q}_2)} + 
    \frac{\cosh{2 \hat{p}_2} }{\hat{q}_2 (\hat{q}_2 - \hat{q}_1) (\hat{q}_2 + \hat{q}_1)}.
\end{equation}
The second Hamiltonian after the change of variables takes the form
\begin{equation}
    \hat{H}_2 = \frac{1}{2} m_2(g g^\dagger) = 
        \frac{\cosh(2\hat{p}_1 + 2 \hat{p}_2)}
            {4\hat{q}_1 \hat{q}_2 (\hat{q}_1 + \hat{q}_2)^2} + 
        \frac{\cosh(2\hat{p}_1 - 2\hat{p}_2)}
            {4 \hat{q}_1 \hat{q}_2 (\hat{q}_1 - \hat{q}_2)^2} + 
        \frac{1}{(\hat{q}^2_1 - \hat{q}^2_2)^2}.
\end{equation}

\newpage

\begin{small}

\end{small}


\begin{thebibliography}{99}
\addcontentsline{toc}{section}{References}

\footnotesize{

\bibitem{AMM} 
A. Alekseev, A. Malkin, and E. Meinrenken, 
\textit{ Lie group valued momentum maps}, 
J. Differential Geom. 48 (1998), no. 3, 445–495

\bibitem{B} 
O. Babelon, 
\textit{ Equations in Dual Variables for Whittaker Functions}, 
Letters in Mathematical Physics, Volume 65, pages 229–240, (2003)

\bibitem{BMMM} 
H.W. Braden, A. Marshakov, A. Mironov, A. Morozov, 
\textit{ On Double-Elliptic Integrable Systems. 1. A Duality Argument for the case of SU(2)}, 
Nuclear Physics B Volume 573, Issues 1–2, 1 May 2000, Pages 553-572

\bibitem{C} 
F. Calogero, 
\textit{ Motion of poles and zeros of special solutions of nonlinear and linear partial differential equations
and related ‘solvable’ many body problems}, 
Nuovo Cimento B, 43, 177–241 (1978)

\bibitem{DE} 
J.F. van Diejen, E. Emsiz, 
\textit{ Bispectral dual difference equations for the quantum Toda chain with boundary perturbations}, 
International Mathematics Research Notices 2019, No. 12, 3740--3767

\bibitem{E} 
P. Etingof, 
\textit{ Whittaker functions on quantum groups and q-deformed Toda operators}, 
Amer. Math. Soc. Transl. Ser.2, vol. 194, 9-25, Amer.Math.Soc., Providence, Rhode Island, 1999 

\bibitem{F} 
L. Feher, 
\textit{ Action-angle map and duality for the open Toda lattice in the perspective of Hamiltonian reduction},
Phys. Lett. A 377 (2013) 2917-2921.

\bibitem{FA} 
L. Feh\'er, V. Ayadi, 
\textit{ Trigonometric Sutherland systems and their Ruijsenaars duals from symplectic reduction}, J
. Math. Phys. 51 (2010) 103511

\bibitem{FG}
L. Feher, T. F. Gorbe, \textit{Duality between the trigonometric BC(n) Sutherland system and a completed rational Ruijsenaars-Schneider-van Diejen system}, J. Math. Phys. 55:102704, 2014 

\bibitem{FK}
L. Feh\'er, C. Klimcik, 
\textit{ On the duality between the hyperbolic Sutherland and the rational Ruijsenaars--Schneider models}, 
J. Phys. A: Math. Theor. 42 (2009) 185202

\bibitem{FK1}
L. Feh\'er, C. Klimcik, 
\textit{ Poisson-Lie interpretation of trigonometric Ruijsenaars duality}, 
Commun. Math. Phys. 301 (2011) 55-104

\bibitem{FK2}
L. Feh\'er and C. Klimcik, 
\textit{ Self-duality of the compactified Ruijsenaars--Schneider system from quasi-Hamiltonian reduction}, 
Nucl. Phys. B 860 (2012) 464-515

\bibitem{FK3}
L. Feh\'er and T.J. Kluck, 
\textit{ New compact forms of the trigonometric Ruijsenaars--Schneider system}, 
Nucl. Phys. B 882 (2014) 97-127

\bibitem{FMarshal}
L. Feh\'er, I. Marshall, 
\textit{ The action-angle dual of an integrable Hamiltonian system of Ruijsenaars--Schneider--van Diejen type}, 
J. Phys. A: Math. Theor. 50 (2017) 314004

\bibitem{FP}
L. Feh\'er and B.G. Pusztai, 
\textit{ A class of Calogero type reductions of free motion on a simple Lie group}, 
Lett. Math. Phys. 79 (2007) 263-277

\bibitem{FT} 
M. Finkelberg, A. Tsymbaliuk, 
\textit{ Multiplicative slices, relativistic Toda and shifted quantum affine algebras}, 
Representations and Nilpotent Orbits of Lie Algebraic Systems, 2019

\bibitem{FGNR} 
V. Fock, A. Gorsky, N. Nekrasov and V. Rubtsov, 
\textit{ Duality in Integrable Systems and Gauge Theories}, 
JHEP 0007 (2000) 028

\bibitem{FM} 
V. Fock, A. Marshakov, 
\textit{ A note on quantum groups and relativistic Toda theory}, 
Nuclear Physics B - Proceedings Supplements, Volume 56, Issue 3, July 1997, Pages 208-214



\bibitem{GMMMO}
A. Gerasimov, A. Marshakov, A. Mironov, A. Morozov, A. Orlov, 
\textit{ Matrix models of two-dimensional gravity and Toda theory}, 
Nucl. Phys. B 357 (1991) 565

\bibitem{GK} 
A. Givental, B. Kim, 
\textit{ Quantum cohomology of flag manifolds and Toda lattices}, 
Commun.Math.Phys. 168 (1995) 609-642

\bibitem{G1}
A. Gorsky, 
\textit{ The Toda system and solution to the N = 2 SUSY Yang--Mills theory}, 
Journal of Physics A: Mathematical and Theoretical, Volume 51, Number 30, 2018

\bibitem{GKMMM}
A. Gorsky, I. Krichever, A. Marshakov, A. Mironov, A. Morozov, 
\textit{ Integrability and Seiberg-Witten Exact Solution}, 
Phys.Lett. B355 (1995) 466-474

\bibitem{GVZ} 
A. Gorsky, M. Vasilyev, A. Zotov, 
\textit{ Dualities in quantum integrable many-body systems and integrable probabilities - I}, 
J. High Energ. Phys. 2022, 159 (2022)

\bibitem{I} 
V.I. Inozemtsev, 
\textit{ Finite Toda lattice}, 
Comm. Math. Phys. 121 (1989), 629–638

\bibitem{Iwasawa}
K. Iwasawa, \textit{On Some Types of Topological Groups}, Annals of Mathematics, Second Series, Vol. 50, No. 3 (Jul., 1949), pp. 507-558

\bibitem{KKS} 
D. Kazhdan, B, Kostant, S. Sternberg, 
\textit{ Hamiltonian group actions and dynamical systems of Calogero type}, 
Comm. Pure Appl. Math. 31 (1978) 481–507

\bibitem{K} 
B. Kostant, 
\textit{ The Solution to a generalized Toda lattice and representation theory}, 
Adv. Math. 34 (1979) 

\bibitem{K2} 
B. Kostant, 
\textit{ On Whittaker vectors and representation theory}, 
Inventiones Math., 48, 1978, 101-184

\bibitem{K1} 
B. Kostant, 
\textit{ Quantization and representation theory}, 
Representation Theory of Lie Groups, Proc. of Symp., Oxford, 1977, pp. 287-317, 
London Math. Soc. Lecture Notes series, 34, Cambridge, 1979

\bibitem{Lu} 
J-H. Lu,  
\textit{ Momentum mapping and reductions of Poisson action}, 
Symplectic geometry, groupoids and Integrable systems, pp. 209-226, (Berkeley, CA, 1989)

\bibitem{MW} 
J. Marsden, A. Weinstein, 
\textit{ Reduction of symplectic manifolds with symmetry}, 
Rep. Math. Phys. 5 (1974) 121-130

\bibitem{Mars} 
A. Marshakov, 
\textit{ Duality in integrable systems and generating functions for new Hamiltonians}, 
Phys.Lett. B476 (2000) 420-426

\bibitem{MM} 
A. Mironov, A. Morozov, 
\textit{ On the status of DELL systems}, 
Nuclear Physics B Volume 999, February 2024, 116448

\bibitem{O} 
M. Olshanetsky, 
\textit{ Solutions of the periodic Toda lattice via the projection procedure and by the algebra-geometric method}, 
Theoretical and Mathematical Physics, Vol. 128, No. 3, pp. 1225–1235, 2001

\bibitem{OP} 
M. Olshanetsky, A. Perelomov, 
\textit{ Quantum integrable systems related to Lie algebras}, 
Physics Reports Volume 94, Issue 6, March 1983, Pages 313-404

\bibitem{Perelomov_lectures} 
A.M. Perelomov, 
\textit{ Integrable Systems of Classical Mechanics and Lie Algebras}, 
Birkh\"auser, 1990

\bibitem{P} 
A. Pogrebkov, 
\textit{ Multiplicative dynamical systems in terms of the induced dynamics}, 
Theoretical and Mathematical Physics, Volume 204, pages 1201-1208, 2020

\bibitem{R1}
S. Ruijsenaars, 
\textit{ Action-angle maps and scattering theory for some finite-dimensional
integrable systems I. The pure soliton case}, 
Commun. Math. Phys. 115, 127-165 (1988)

\bibitem{R2}
S. Ruijsenaars, 
\textit{ Action-angle maps and scattering theory for some finite-dimensional
integrable systems II. Solitons, antisolitons and their bound states}, 
Publ. RIMS 30, 865-1008 (1994)

\bibitem{R4}
S. Ruijsenaars, 
\textit{ Finite-dimensional soliton systems}, 
In: Kupershmidt, B. (ed.) Integrable and Superintegrable Systems, pp. 165–206. 
World Scientific, Singapore (1990)

\bibitem{R3}
S. Ruijsenaars, 
\textit{ Action-angle maps and scattering theory for some finite-dimensional
integrable systems III. Sutherland type systems and their duals}, 
Publ. RIMS 31, 247-353 (1995)

\bibitem{R} 
S. Ruijsenaars, 
\textit{ Relativistic Toda systems}, 
Commun. Math. Phys. 133, 217-247 (1990) 
 
\bibitem{RS} 
S. Ruijsenaars, H. Schneider, 
\textit{ A new class of integrable systems and its relation to solitons}, 
Annals of Physics 146 (1986) 1–34

\bibitem{SK} 
E. Sklyanin, 
\textit{ Bispectrality for the quantum open Toda chain}, 
J. Phys. A: Math and Theor, 46:38 (2013) 382001

\bibitem{T}
M. Toda, 
\textit{ Theory of nonlinear lattices}, 
Berlin, Heidelberg, New York: Springer 1981

\bibitem{Z} 
A. Zabrodin, 
\textit{ Elliptic solutions to integrable nonlinear equations and many-body systems}, 
J. Geom. Phys., 146 (2019) 103506

\bibitem{ZLB} 
Z. Zakirova, V.Lunev, N. Beloborodov, 
\textit{ pq-Duality: a set of simple examples}, 
Methods of theoretical physics, Volume 118, pages 142–145, (2023)

%


}
\end{thebibliography}
\end{document}